\documentstyle[12pt]{article}
\newcommand{\doublespace}{\renewcommand{\baselinestretch}{1.75}
\Large\normalsize}

\begin{document}
\doublespace
\begin{titlepage}

\centerline{\bf Gravitational energy of rotating black holes}
\bigskip
\centerline{\it J. W. Maluf$\,^{*}$, E. F. Martins and A. Kneip}
\centerline{\it Departamento de F\'isica}
\centerline{\it Universidade de Bras\'ilia}
\centerline{\it C.P. 04385}
\centerline{\it 70.919-970  Bras\'ilia, DF}  
\centerline{\it Brazil}
\date{}
\begin{abstract}
In the teleparallel equivalent of general relativity the energy density
of asymptoticaly flat gravitational fields can be naturally defined
as a scalar density restricted to a three dimensional spacelike
hypersurface $\Sigma$. Integration over the 
whole $\Sigma$ yields the standard ADM energy. Here we obtain the 
formal expression of the localized energy for a Kerr black hole.
The expression  of the energy inside a surface of constant radius
can be explicitly calculated in the limit of small $a$, the
specific angular momentum. Such expression turns out to be exactly
the same as the one obtained by means of the method proposed 
recently by Brown and York. We also calculate the energy contained
within the outer horizon of the black hole, for {\it any} value of 
$a$. The result is practically indistinguishable from $E=2M_{ir}$, 
where $M_{ir}$ is the irreducible mass of the black 
hole.
\end{abstract}
\thispagestyle{empty}
\vfill
\par
\noindent (*) e-mail: wadih@guarany.cpd.unb.br
\end{titlepage}
\newpage

\noindent {\bf I. Introduction}\par
\bigskip
Although it is widely believed that Einstein's 
equations describe the dynamics of the gravitational field, it 
has not been possible so far to arrive at a definite expression
for the gravitational energy in the context of Einstein's general
relativity. Attempts based on the 
Hilbert-Einstein action integral fail to yield an expression 
for the gravitational energy {\it density}\cite{Landau,MTW}.
The {\it total} gravitational energy is normally obtained
from surface terms in the action or in the 
Hamiltonian\cite{ADM,Regge},
or from pseudotensor methods which make use of coordinate
dependent expressions.

Recently an expression for quasi-local energy has been
proposed by Brown and York\cite{Brown}. Such expression is derived
directly from the action functional $A_{cl}$. The latter is identified 
as Hamilton's principal function and, in similarity  with the classical
Hamilton-Jacobi equation, which expresses the energy of a classical
solution as minus the time rate of the change of the action, the 
quasilocal gravitational energy is identified as minus the proper
time rate of change of the Hilbert-Einstein action (with surface terms
included). Expressions for the quasilocal energy have been obtained
for the Schwarzschild solution\cite{Brown} and for the Kerr 
solution\cite{Martinez}. 

Einstein's equations can also be obtained from the teleparallel 
equivalent of general relativity (TEGR). The Lagrangian formulation
of the TEGR is established by means of the tetrad field 
$e^a\,_\mu$ and the spin affine connection $\omega_{\mu ab}$, which
are taken to be completely independent field variables, even at the
level of field equations. This formulation has been investigated
in the past in the context of Poincar\'e gauge 
theories\cite{Hehl1,Hehl2}. However, as we will explain ahead, 
this is not an alternative theory of gravity. This is just an
{\it alternative formulation} of general relativity, in which 
the curvature tensor constructed out of $\omega_{\mu ab}$ vanishes,
but the torsion tensor is non-vanishing. The physical content of
the theory is dictated by Einsten's equations.  In this
alternative geometrical formulation the gravitational energy
density can be naturally defined.

The expression for the gravitational energy density arises in the
framework of the Hamiltonian formulation of the TEGR\cite{Maluf1}.
It has been demonstrated that under a suitable gauge fixing of
$\omega_{\mu ab}$, already at the Lagrangian level, the Hamiltonian
formulation of the TEGR is well defined. The resulting constraints
are first class constraints\cite{Maluf1}. The Hamiltonian 
formulation turns out to be very much similar  to the usual ADM 
formulation\cite{ADM}. However there are  crucial differences.
The integral form of the Hamiltonian constraint equation
$C=0$ in the TEGR can be written in the form $C=H-E_{ADM}=0$, when
we restrict considerations to asymptotically flat 
spacetimes\cite{Maluf2}. The quantity $\varepsilon(x)$ which appears
in the expression of $C$ and which under integration yields
$E_{ADM}$ is recognized as the gravitational energy density.

We have calculated the energy inside a sphere of radius $r_o$ in a
Schwarzschild spacetime by means of $\varepsilon(x)$\cite{Maluf2}.
The expression turns out to be exactly the same as the one obtained 
by means of the procedure of ref.\cite{Brown} (expression (6.14) of
\cite{Brown}). In this paper we consider the Kerr black-hole.
We obtain the formal expression for the energy contained in any
space volume in terms of non-trivial integrals. In the 
limit of slow rotation (small specific angular momentum) the energy
contained within a surface of constant radius $r_o$ 
can be calculated. Again the result obtained here is exactly
the same as that obtained by Martinez\cite{Martinez} who adopted
Brown and York's procedure. The advantage of our procedure rests on
the fact that the localized energy associated with a Kerr
spacetime can be calculated in the general case, without recourse
to particular limits, at least by means of numerical integration, 
whereas in Brown and York's procedure one has to calculate the
subtraction term $\varepsilon^0$ and for this purpose 
it is necessary to 
embed an arbitrary two dimensional boundary surface of the Kerr
space $\Sigma$ in the appropriate reference space ($E^3$, say), 
which is not always possible\cite{Martinez}.

We have also calculated the energy contained within the outer
horizon of the black hole. Such a quantity has been obtained 
by Martinez\cite{Martinez} in the limit of small $a$, and 
reads $E=2M_{ir}$ (plus corrections of order 
$O({{a^4}\over {M_{ir}^4}})\,$), where $M_{ir}$ is the irreducible mass
of the black hole. The concept of irreducible mass was 
introduced by Christodoulou\cite{Christodoulou}. He showed that
the mass of a rotating black hole cannot be decreased to values
below $M_{ir}$  by means of Penrose's process of extraction 
of energy. One would thus consider $E=2M_{ir}$
to be the energy that cannot escape from the black hole. 
Here we obtain the expression of the energy contained within the 
horizon for  {\it any} value of $a$. The result is striking.
The numerical values of this
expression are practically coincident with $2M_{ir}$ in the
whole range $0\le a \le m$, although the expression is 
is algebrically different from $2M_{ir}$.

In section II we present the mathematical preliminaries of the TEGR,
its Hamiltonian formulation and the expression of the energy
for an arbitrary asymptoticaly flat
spacetime. In section III we carry out the
construction of triads for a three dimensional spacelike
hypersurface of the Kerr type, obtain the general expression of 
the energy contained in a volume $V$ of space and provide the 
exact expression of the latter  in the limit of slow rotation.
Comments and conclusions are presented on section IV.

Notation: spacetime indices $\mu, \nu, ...$ and local Lorentz indices
$a, b, ...$ run from 0 to 3. In the 3+1 decomposition latin indices 
from the middle of the alphabet indicate space indices according to
$\mu=0,i,\;\;a=(0),(i)$. The tetrad field $e^a\,_\mu$ and
the spin connection $\omega_{\mu ab}$ yield the usual definitions
of the torsion and curvature tensors:  $R^a\,_{b \mu \nu}=
\partial_\mu \omega_\nu\,^a\,_b +
\omega_\mu\,^a\,_c\omega_\nu\,^c\,_b\,-\,...$,
$T^a\,_{\mu \nu}=\partial_\mu e^a\,_\nu+
\omega_\mu\,^a\,_b\,e^b\,_\nu\,-\,...$. The flat spacetime metric 
is fixed by $\eta_{(0)(0)}=-1$. \\

\bigskip
\bigskip
\noindent {\bf II. The TEGR in Hamiltonian form}\par
\bigskip
In the TEGR the tetrad field $e^a\,_\mu$ and the spin connection
$\omega_{\mu ab}$ are completely independent field variables. The 
latter is enforced to satisfy the condition of zero curvature. 
The Lagrangian density in empty spacetime is given by

$$L(e,\omega,\lambda)\;=\;-ke({1\over 4}T^{abc}T_{abc}\,+\,
{1\over 2}T^{abc}T_{bac}\,-\,T^aT_a)\;+\;
e\lambda^{ab\mu\nu}R_{ab\mu\nu}(\omega)\;.\eqno(1)$$

\noindent where $k={1\over {16\pi G}}$, $G$ is the gravitational 
constant; $e\,=\,det(e^a\,_\mu)$, $\lambda^{ab\mu\nu}$ are 
Lagrange multipliers and $T_a$ is the trace of the torsion tensor
defined by $T_a=T^b\,_{ba}$.   

The equivalence of the TEGR with Einstein's general relativity is         
based on the identity

$$eR(e,\omega)\;=\;eR(e)\,+\,
e({1\over 4}T^{abc}T_{abc}\,+\,T^{abc}T_{acb}\,-\,T^aT_a)\,-\,
2\partial_\mu(eT^{\mu})\;,\eqno(2)$$

\noindent which is obtained by just substituting the arbitrary
spin connection $\omega_{\mu ab}\,=\,^o\omega_{\mu ab}(e)\,+\,
K_{\mu ab}$ in the scalar curvature tensor $R(e,\omega)$ in the
left hand side; $^o\omega_{\mu ab}(e)$ is the Levi-Civita 
connection and $K_{\mu ab}\,=\,
{1\over 2}e_a\,^\lambda e_b\,^\nu(T_{\lambda \mu \nu}+
T_{\nu \lambda \mu}-T_{\mu \nu \lambda})$ is the contorsion tensor.
The vanishing of $R^a\,_{b\mu\nu}(\omega)$, which is one of the
field equations derived from (1), implies the equivalence of 
the scalar curvature $R(e)$, constructed out of $e^a\,_\mu$ only, 
and the quadratic combination of the torsion tensor. It also
ensures that the field equation arising from the variation of
$L$ with respect to $e^a\,_\mu$ is strictly equivalent to
Einstein's equations in tetrad form. Let 
${{\delta L}\over {\delta {e^{a\mu}}}}=0$ denote the field equations
satisfied by $e^{a\mu}$. It can be shown by explicit calculations
that

$${{\delta L}\over {\delta {e^{a\mu}}}}\;=\;
{1\over 2}\lbrace R_{a\mu}(e)\,
-\,{1\over 2}e_{a\mu}R(e)\rbrace\;.\eqno(3)$$

(we refer the reader to
ref.\cite{Maluf1} for additional details).

It is important to notice that for asymptotically flat spacetimes
the total divergence in (2) does {\it not} contribute to the 
action integral. This term is a scalar density that falls off as 
$1\over {r^3}$ when $r \rightarrow \infty$. In this limit we should
consider variations in $g_{\mu \nu}$ or in $e_{a\mu}$ that preserve
the asymptotic structure of the flat spacetime metric; the allowed
coordinate transformations must be of the Poincar\'e type. 
The variation of $\partial_\mu (eT^\mu)$ at infinity
under such variations
of $e_{a\mu}$  vanishes. Moreover all surface integrals arising from
partial integration in the variation of the action integral vanish
as well. Therefore the action does not require additional surface
terms, as it is invariant under transformations that preserve the
asymptotic structure of the field quantities. This property fixes
the action integral, together with the requirement that the variation
of the latter must yield Einstein's equations (the Hilbert-Einstein 
Lagrangian requires the addition of a surface term for the variation
of the action 
to be well defined; a clear discussion of this point is given in 
ref.\cite{Faddeev}). In what follows 
we will be interested in asymptoticaly flat spacetimes.

The Hamiltonian formulation of the TEGR can be successfully 
implemented if we fix the gauge $\omega_{0ab}=0$ from the 
outset, since in this case the constraints (to be 
shown below) constitute a {\it first class} set\cite{Maluf1}.
The condition $\omega_{0ab}=0$ is achieved by breaking the local
Lorentz symmetry of (1). We still make use of the residual time
independent gauge symmetry to fix the usual time gauge condition
$e_{(k)}\,^0\,=\,e_{(0)i}\,=\,0$. Because of $\omega_{0ab}=0$,
$H$ does not depend on $P^{kab}$, the momentum canonically 
conjugated to $\omega_{kab}$. Therefore arbitrary variations of
$L=p\dot q -H$ with respect to $P^{kab}$ yields 
$\dot \omega_{kab}=0$. Thus in view of $\omega_{0ab}=0$, 
$\omega_{kab}$ drops out from our considerations. The above 
gauge fixing can be understood as the fixation of a {\it global}
reference frame.    

Under the above gauge fixing the canonical action integral obtained
from (1) becomes\cite{Maluf1}

$$A_{TL}\;=\;\int d^4x\lbrace \Pi^{(j)k}\dot e_{(j)k}\,-\,H\rbrace\;,
\eqno(4)$$

$$H\;=\;NC\,+\,N^iC_i\,+\,\Sigma_{mn}\Pi^{mn}\,+\,
{1\over {8\pi G}}\partial_k (NeT^k)\,+\,
\partial_k (\Pi^{jk}N_j)\;.\eqno(5)$$

\noindent $N$ and $N^i$ are the lapse and shift functions, and 
$\Sigma_{mn}=-\Sigma_{nm}$ are Lagrange multipliers. The constraints
are defined by 

$$ C\;=\;\partial_j(2keT^j)\,-\,ke\Sigma^{kij}T_{kij}\,-\,
{1\over {4ke}}(\Pi^{ij}\Pi_{ji}-{1\over 2}\Pi^2)\;,\eqno(6)$$

$$C_k\;=\;-e_{(j)k}\partial_i\Pi^{(j)i}\,-\,
\Pi^{(j)i}T_{(j)ik}\;,\eqno(7)$$

\noindent with $e=det(e_{(j)k})$ and $T^i\,=\,g^{ik}e^{(j)l}T_{(j)lk}$. 
We remark that (4) and (5) are invariant under global SO(3) and
general coordinate transformations.  

We assume the asymptotic behaviour $e_{(j)k}\approx \eta_{jk}+
{1\over 2}h_{jk}({1\over r})$ for $r \rightarrow \infty$. In view
of the relation

$${1\over {8\pi G}}\int d^3x\partial_j(eT^j)\;=\;
{1\over {16\pi G}}\int_S dS_k(\partial_ih_{ik}-\partial_kh_{ii})
\; \equiv \; E_{ADM}\;\eqno(8)$$

\noindent where the surface integral is evaluated for 
$r \rightarrow \infty$, the integral form of 
the Hamiltonian constraint $C=0$ may be rewritten as

$$\int d^3x\biggl\{ ke\Sigma^{kij}T_{kij}+
{1\over {4ke}}(\Pi^{ij}\Pi_{ji}-{1\over 2}\Pi^2)\biggr\}
\;=\;E_{ADM}\;.\eqno(9)$$

\noindent The integration is over the whole three dimensional
space. Given that $\partial_j(eT^j)$ is a scalar  density,
from (7) and (8) we define the gravitational
energy density enclosed by a volume V of the space as

$$E_g\;=\;{1\over {8\pi G}}\int_V d^3x\partial_j(eT^j)\;.\eqno(10)$$  

\noindent It must be noted that $E_g$ depends only on the triads
$e_{(k)i}$ restricted to a three-dimensional spacelike hypersurface;
the inverse quantities $e^{(k)i}$ can be written in terms of 
$e_{(k)i}$. From the identity (3) we observe that the dynamics of
the triads does not depend on $\omega_{\mu ab}$. Therefore $E_g$
given above does not depend on the fixation of any gauge for
$\omega_{\mu ab}$.   \\

\bigskip
\bigskip

\noindent {\bf III. Energy of the Kerr geometry}\par
\bigskip
The Kerr solution\cite{Kerr} describes the field of a 
rotating black hole. In terms of Boyer and Lindquist
coordinates\cite{Boyer}  $(t,r,\theta,\phi)$ it is described 
by the metric

$$ds^2\;=\;-{\Delta \over {\rho^2}}\lbrack dt-
a\,sin^2\theta d\phi\rbrack^2\;+\;
{{sin^2\theta}\over {\rho^2}}\lbrack(r^2 + a^2)d\phi-a\,dt\rbrack^2
\;+\;{{\rho^2}\over \Delta}dr^2\;+\;\rho^2d\theta^2\;,\eqno(11)$$

$$\Delta\; \equiv \;r^2-2mr+a^2\;,$$

$$\rho^2 \; \equiv \; r^2+a^2\,cos^2\theta\;;$$

\noindent $a$ is the specifc angular momentum defined by
$a\,=\,{J\over m}$. The components of the metric restricted to 
the three dimensional spacelike hypersurface are given by
$g_{11}={{\rho^2}\over \Delta}$, $g_{22}=\rho^2$ and
$g_{33}={{\Sigma^2}\over {\rho^2}}sin^2\theta$, where $\Sigma$ is
defined by

$$\Sigma^2\;=\;(r^2+a^2)^2\,-\,\Delta\,a^2\,sin^2\theta\;.$$  

We define the triads $e_{(k)i}$ as

$$e_{(k)i}\;=\;
\pmatrix{  {\rho \over \sqrt{\Delta}}sin\theta\,cos\phi &
\rho cos\theta\,cos\phi & 
-{\Sigma \over \rho}sin\theta\,sin\phi \cr
{\rho \over \sqrt{\Delta}}sin\theta\,sin\phi &
\rho cos\theta\,sin\phi &
{\Sigma \over \rho}sin\theta\,cos\phi \cr
{\rho \over \sqrt{\Delta}}cos\theta &
-\rho sin\theta& 0 \cr } \eqno(12)$$  

\noindent $(k)$ is the line index and $i$ is the column index. The
one form $e^{(k)}$ is defined by

$$e^{(k)}\;=\;e^{(k)}\,_rdr\,+\,e^{(k)}\,_\theta d\theta\,+\,
e^{(k)}\,_\phi d\phi\;,$$

\noindent from what follows

$$e^{(k)}e_{(k)}\;=\;{\rho^2 \over \Delta}dr^2\,+\,
\rho^2 d\theta^2\,+\,{\Sigma^2 \over \rho^2}sin^2\theta d\phi^2$$

\noindent We also obtain $e=det(e_{(k)i})=
{{\rho \Sigma }\over \sqrt{\Delta}}sin\theta$. Therefore the triads
given by (12) describe the components of the Kerr solution 
restricted to the three dimensional spacelike hypersurface.

One readily notices that there is another set of triads that yields
the Kerr solution, namely, the set which is diagonal and whose
entries are given by the square roots of $g_{ii}$. This set is
not appropriate for our purposes, and the reason can be
understood even in the simple clase of flat spacetime.
In the limit when both $a$ and $m$ go to zero (12) describes
flat space: the curvature tensor {\it and} the torsion tensor
vanish in this case. However, for the diagonal set of triads
(again requiring $a\rightarrow 0$ and $m \rightarrow 0$), 

$$e^{(r)}=dr\;,\;e^{(\theta)}=r\,d\theta\;,\;
e^{(\phi)}=r\,sin\theta\, d\phi\;,$$

\noindent some components of the torsion tensor do not vanish,
$T_{(2)12}=1$, $T_{(3)13}=sin\theta$, and $E_g$ calculated out 
of the diagonal set above diverges when integrated over 
the whole space. Therefore the use of (12) is mandatory in the
present context.

The components of the torsion tensor can be calculated in a
straightforward way from (12). Only $T_{(3)13}$ and $T_{(3)23}$ 
are vanishing. The others are given by:

$$T_{(1)12}\;=\;cos\theta cos\phi\,({r \over \rho}+
{a^2 \over {\rho \sqrt{\Delta}}}\,sin^2\theta - 
{\rho \over {\sqrt{\Delta}}})$$

$$T_{(1)13}\;=\;sin\theta sin\phi 
\lbrace -{1\over {\rho \Sigma}}[2r(r^2+a^2)-a^2 sin^2\theta\,(r-m)]+
{{r\Sigma}\over \rho^3}+{\rho \over {\sqrt{\Delta}}}\rbrace$$

$$T_{(1)23}\;=\;cos\theta sin\phi \, \lbrace
\rho-{\Sigma \over \rho}+
a^2 sin^2\theta\,({\Delta \over {\rho \Sigma}}-{\Sigma \over \rho^3})
\rbrace$$

$$T_{(2)12}\;=\;cos\theta sin\phi\,({r \over \rho}+
{a^2 \over {\rho\sqrt{\Delta}}}\,sin^2\theta-
{\rho \over \sqrt{\Delta}})$$

$$T_{(2)13}\;=\;-sin\theta cos\phi\, \lbrace
-{1\over {\rho \Sigma}}[2r(r^2+a^2)-
a^2 sin^2\theta\,(r-m)]+
{{r\Sigma} \over \rho^3}+{\rho \over \sqrt{\Delta}} \rbrace$$

$$T_{(2)23}\;=\;-cos\theta cos\phi\,\lbrace \rho -
{\Sigma \over \rho}+a^2\sin^2\theta\,({\Delta \over {\rho \Sigma}}-
{\Sigma\over \rho^3}) \rbrace$$

$$T_{(3)12}\;=\;sin\theta\,[-{r \over \rho}+
{\rho \over \sqrt{\Delta}}+
{a^2 \over {\rho\sqrt{\Delta}}}\,cos^2\theta]$$

In order to evaluate (9) we need to obtain $T^i$.
After a long calculation we arrive at

$$T^1\;=\;{\sqrt{\Delta} \over {\rho^2}}\;+\;
{\sqrt{\Delta} \over \Sigma}\;-\;
{\Delta \over {\rho^2 \Sigma^2}}\lbrack 2r(r^2+a^2)\,-\,
a^2sin^2\theta(r-m)\rbrack \;,$$

$$T^2\;=\;sin\theta\,cos\theta\,{a^2 \over {\rho^4}}\,+\,
{1\over{\rho\Sigma}}\,{{cos\theta}\over{\sin\theta}}\,\lbrack\rho
-{\Sigma\over \rho}+a^2sin^2\theta({\Delta\over{\rho\Sigma}}-
{\Sigma\over{\rho^3}})\rbrack\;,$$

$$T^3\;=\;0\;.$$  

The gravitational energy density inside a volume $V$ of a three
dimensional spacelike hypersurface of the Kerr solution can now 
be easily calculated. It is given by

$$E_g\;=\;{1\over {8\pi}}\int_V dr\,d\theta\,d\phi 
\biggl\{ {\partial \over{\partial r}}\biggl[ sin\theta \lbrack \rho+
{\Sigma \over \rho}-{\sqrt{\Delta} \over{\rho\Sigma}}\,
\biggl(  \;2r(r^2+
a^2)-a^2sin^2\theta(r-m)\; \biggr)\rbrack \biggr]$$

$$+{\partial\over{\partial \theta}}\biggl[
{{\Sigma a^2}\over{\sqrt{\Delta}\rho^3}}\,sin^2\theta cos\theta+
{cos\theta \over \sqrt{\Delta}}
\biggl(\;\rho-{\Sigma \over \rho}+a^2sin^2\theta(
{\Delta\over{\rho\Sigma}}-{\Sigma\over{\rho^3}}\,)\;\;
\biggr)\biggr]  \biggr\} \eqno(13) $$ 

Next we specialize $E_g$ to the case when the volume $V$ is 
contained within a surface with constant radius $r=r_o$ assuming
$r_o \ge r_+$, where $r_+\,=\,m+\sqrt{m^2-a^2}$ is the outer
horizon of the black hole. The
integrations in $\phi$ and $r$ are trivial. Also, because
we integrate $\theta$ between 0 and $\pi$, the second line
of the expression above vanishes. We then obtain

$$E_g\;=\;{1\over 4}\int_0^\pi d\theta\,sin\theta \biggl\{
\rho+{\Sigma \over \rho} - 
{\sqrt{\Delta} \over {\rho \Sigma}}\,\biggl( 2r(r^2+a^2)-
a^2sin^2\theta
(r-m)\, \biggr) \biggr\}_{r=r_o}\;.\eqno(14)$$

We have not managed to evaluate exactly the integral above.
However, in the limit of slow rotation, namely, when 
${a\over r_o}\,<<\,1$ all integrals have a simple
structure and we can obtain the  
approximate expression of $E_g$. It reads

$$E_g\;=\;r_o\biggl( 1-\sqrt{1-{{2m} \over r_o}+
{a^2\over r_o^2}}\biggr) 
\;+\; {a^2\over {6r_o}}\,\biggl[ 2+{{2m} \over r_o}+
\biggl( 1+{{2m}\over r_o}\biggr)
\sqrt{1-{{2m} \over r_o}+{a^2\over r_o^2}}
\biggr]\eqno(15)$$

\noindent This is exactly the expression found by 
Martinez\cite{Martinez} for the energy inside the surface of
constant radius $r_o$ in a spacelike hypersurface of a Kerr
black hole, in the limit of small specific angular momentum.
As in ref.\cite{Martinez}, we have not expanded the square root
which appears in (15) in powers in $a^2 \over r_o^2$.

We remark that the expansion of $\rho+
{\Sigma \over \rho}$ in the integrand of (14) 
yields $-\varepsilon_0$,
whereas the remaining term corresponds exactly to
$\varepsilon$, expressions (3.17) and (3.1) respectively
of \cite{Martinez}.  It does not seem to 
be possible,      
however, to split $\partial_i(eT^i)$ into two terms such that
their integrals arise in the form $\varepsilon - 
\varepsilon_0$.  

As a very interesting application of (14), let us calculate
the energy contained within the outer horizon, i.e., we 
will calculate (14) when the surface of constant radius 
is defined by $r_o=r_+$. This surface is characterized by
$\Delta=0$. The integral can be calculated exactly for
{\it any} value of $a$. The latter is parametrized in terms
of the black hole mass $m$ according to

$$a\;=\;km\;\;\;,\;\;\;0\le k \le 1\;.$$

\noindent After a number of integrations we arrive at

$$E_g\;=\;m\biggl[ {\sqrt{2p}\over 4}+\,
\,{{6p-k^2}\over 4k}ln \,\biggl(
{{\sqrt{2p} + k}\over p}\biggr) \biggr]\;,\eqno(16)$$

\noindent where $p$ is defined by 

$$p\;=\;1\,+\,\sqrt{1-k^2}\;.$$

\noindent This  is the ammount of energy expected not to escape
from the black hole by any process in which the black hole interacts
with external particles. It is then important to compare (16) with
$E=2M_{ir}$. 

We recall that a rotating black hole can have its
mass decreased by means of Penrose's process of extraction of 
energy\cite{Wald}.
The idea is the following. We consider a particle that is emitted
towards the black hole and penetrates into the ergosphere. Suppose 
we arrange the particle to break up into two fragments, in such a
way that one of the fragments has total negative energy. This is
possible in principle, since the energy need not be positive in
the ergosphere. By conservation of energy, the fragment with
positive energy has an energy greater than that of the incident
particle. Thus energy will be extracted from the black hole if the
positive energy particle escapes to infinity, while the black
hole absorbs the negative energy one. As a consequence, the mass
of the black hole is decreased. We expect, however, that not the 
whole energy of the black hole can be extracted in this manner.
The existence of the horizon certainly prevents one from 
exhausting the total energy. Christodoulou\cite{Christodoulou}
has given an argument to determine how much energy can be extracted
from the black hole by Penrose's process. He concluded that 
at the end of this process (when the ergosphere disappears and
the black hole becomes static)
the final (irreducible) mass of the black hole is given by 

$$M_{ir}\;=\; {1\over 2}\sqrt{r_+^2\,+\,a^2}\;.$$

\noindent Martinez\cite{Martinez} has calculated the energy inside
the horizon of the Kerr black hole in the limit of small $a$.
He arrived at $E=2M_{ir}\lbrack 1+O({a^4 \over M_{ir}^4})\rbrack$.
A similar result (in the same approximation) has been obtained
by Zaslavskii\cite{Zaslavskii} in the analysis of a generic
axially-symmetric spacetime. 
The question immediately arises as to whether this
relationship holds for {\it any} value of $a$. This is in fact 
the conjecture made in ref.\cite{Martinez}.

Since expressions (14) and (16) are valid for {\it any} value of
$a$ in the appropriate range, it is worth comparing (16) with 
$2M_{ir}$. In our parametrization we have

$$E\;=\;2M_{ir}\;=\;m\sqrt{2p}\;.\eqno(17)$$

The expression above certainly looks different from (16).
However in the range $0 \le a \le m$ 
expressions (16) and (17) as functions of $k$ are strikingly
indistinguishable, as we can see in Fig.1. In the latter we have
plotted $E\over m$ against $k$. The upper curve represents (16),
the lower one (17). We see that for small values of the parameter
$k$ the two curves are essentially coincident. A tiny deviation
occurs for values of $k$ near 1. Inspite of this deviation,
this is a remarkable result in favour of (14).

Unfortunately we have not been able to explain such small deviation
between (16) and (17) for values of $k$ near 1, although we expect
such explanation to be of fundamental importance. It might be
related to some physical property of the Kerr black hole which we
do not understand yet.                                    \\

\bigskip
\noindent {\bf Comments}\par
\bigskip

The gravitational energy $E_g$
defined by (14) can be evaluated for an arbitrary value
of $a$ by means of numerical integration. This is the major
advantage of our procedure as compared to that of
Brown and York\cite{Brown}. By means of the latter one
cannot construct expressions like (13) and (14), which may
be useful in the study of astrophysical problems, since 
in a general situation  Brown and York's procedure requires the 
embedding of an arbitrary two dimensional boundary surface 
of the Kerr space in the reference space $E^3$,
a construction which is not possible in general\cite{Martinez}
(the evaluation of $\varepsilon_0$ in \cite{Martinez} is only
possible in the limit ${a\over r_o}\,<<\,1$). Therefore the 
present approach is more general than that of ref.\cite{Martinez}.
Finally we remark that we expect expression (10) to be
useful in the study of the thermodynamics of self-gravitating
systems, where the gravitational energy plays the role of the 
thermodynamical internal energy that is conjugate to the
inverse temperature. We hope to come to this issue in the
future.   \\

\bigskip
\bigskip
\noindent {\it Acknowledgements}\par
\noindent This work was supported in part by CNPQ.

\noindent {\bf Figure Captions}\par
\bigskip
\noindent Figure 1: ${E\over m}$ against $k$. The upper curve
represents the energy expression (16), the lower one (17).

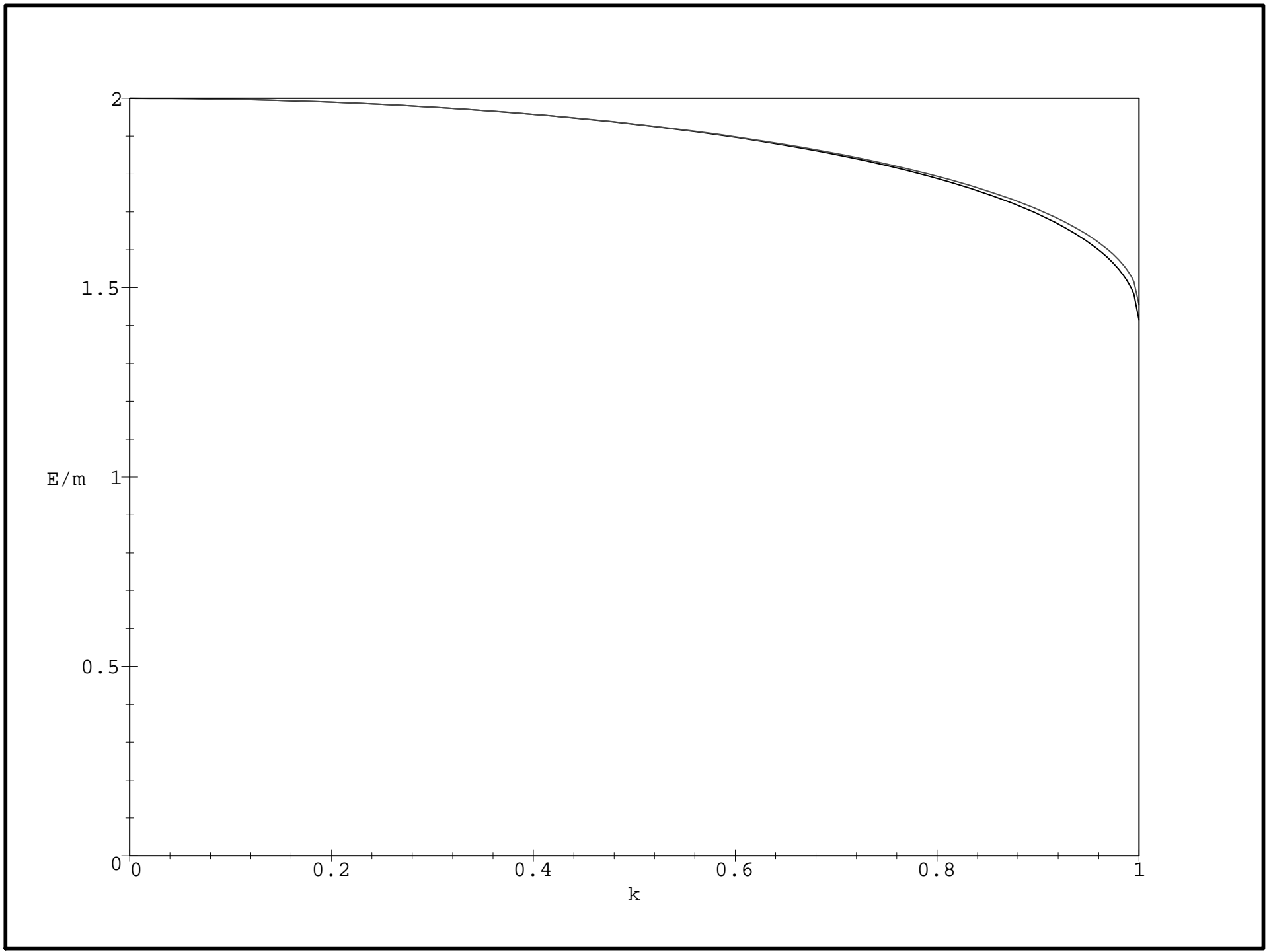
\end{document}